\documentclass[a4paper,UKenglish,cleveref, autoref, thm-restate]{lipics-v2021}

\usepackage{amsmath,amsfonts}
\usepackage{algorithmic}
\usepackage{graphicx}
\usepackage{textcomp}
\usepackage{xcolor}
\usepackage{booktabs}
\usepackage{textcomp}
\usepackage{xcolor}
\usepackage{array}
\usepackage{xspace}
\usepackage{listings}
\usepackage{longtable}
\usepackage{lscape}
\usepackage{threeparttable}
\usepackage{multirow}
\usepackage{float}
\usepackage{placeins}
\usepackage[normalem]{ulem}
\usepackage{pifont}
\usepackage{tcolorbox}

\usepackage{dsfont}
\usepackage{mathtools}

\usepackage{comment}

\usepackage{framed}
\usepackage{hyperref}

\title{Which Neurons Are More Security Sensitive? Probing LLMs to Detect Malicious Code}

\title{Which Neurons Are More Security Sensitive? Probing LLMs' Ability to Detect Malicious Code}

\title{Which Neurons Detect Malicious Code? A Probing Study of LLM Security Knowledge}

\author{Lam D. Dao}{RMIT University Vietnam}{s4019052@rmit.edu.vn}{https://orcid.org/0009-0006-4080-1808}{}

\author{Vang T. Nguyen}{Hanoi University, Vietnam}{12423024@utehy.edu.vn}{https://orcid.org/0009-0006-8379-8645}{}

\author{Anh M. T. Bui}{SOICT, Hanoi University of Science and Technology}{anhbtm@soict.hust.edu.vn}{https://orcid.org/0000-0001-7877-9438}{}

\author{Phuong T. Nguyen}{Universit\`a degli studi dell'Aquila, Italy}{phuong.nguyen@univaq.it}{https://orcid.org/0000-0002-3666-4162}{}

\authorrunning{Lam D. Dao, Vang T. Nguyen, Anh M. T. Bui, and Phuong T. Nguyen} 

\Copyright{Jane Open Access and Joan R. Public} 



\category{} 

\relatedversion{} 

\nolinenumbers 

\EventEditors{John Q. Open and Joan R. Access}
\EventNoEds{2}
\EventLongTitle{The International Symposium on Empirical Software Engineering and Measurement (ESEM 2026)}
\EventShortTitle{ESEM 2026}
\EventAcronym{ESEM}
\EventYear{2016}
\EventDate{December 24--27, 2016}
\EventLocation{Little Whinging, United Kingdom}
\EventLogo{}
\SeriesVolume{42}
\ArticleNo{23}

\newcommand*{\ie}{i.e.,\@\xspace}
\newcommand*{\eg}{e.g.,\@\xspace}

\newcommand*{\Ll}{\texttt{Llama3.1-8\-B-Instruct}\@\xspace}
\newcommand*{\Qw}{\texttt{Qwen2.5-7\-B-Instruct}\@\xspace}
\newcommand*{\Ms}{\texttt{Mistral\-v0.3-7B-Instruct}\@\xspace}

\newcommand{\rqone}{\textbf{RQ$_1$}: \emph{Are the probing mechanisms able to identify the most defensive neurons?}}

\newcommand{\rqtwo}{\textbf{RQ$_2$}: \emph{How are the defensive neurons distributed across the layers and how stable is their selection across runs?}}

\begin{document}

\maketitle

\begin{abstract}
\noindent
\textbf{Background.} Large language models (LLMs) have become increasingly capable of understanding and generating source code, leading to their widespread adoption in software engineering tasks such as code completion, repair, and vulnerability detection. However, despite their strong empirical performance, the internal mechanisms through which LLMs recognize malicious or vulnerable code patterns remain poorly understood.

\noindent
\textbf{Aim.} We investigated where the malware detection behavior is encoded inside LLMs Feed Forward Network (FFN) neurons and 
verified the attribution with causal interventions on the neurons identified. This aims to identify the most important neurons in detecting malicious code.

\noindent
\textbf{Methods.} 
We applied mechanistic interpretability methods to locate the neurons being responsible for malware-detection behavior in three instruction-tuned LLMs: \Ll, \Ms, and \Qw. Using 1,500 malicious and 1,500 benign PyPI packages from the PyPI Malregistry, we attribute the behavior to a set of neurons. 

\noindent
\textbf{Results.} 
The experimental 
results reveal that amplifying facilitating neurons for malware detection while suppressing inhibiting ones can boost accuracy, while the reverse collapses predictions toward a single class, although the magnitude and consistency is heavily model-dependent. We demonstrated that the guardrail detection mechanism varies across models, each represents its malware detection behavior differently within its FFN layers.

\noindent
\textbf{Conclusions.} 
Probing the neurons associated with security-relevant knowledge helps us gain insights into how LLMs encode malicious programming concepts, identify potentially harmful memorized behaviors, paving the way toward more reliable defense mechanisms, such as neuron-level editing, selective unlearning, and security-aware alignment for code-focused LLMs.




\end{abstract}

\section{Introduction}
\label{sec:introduction}

Large Language Models (LLMs) have transformed software engineering (SE) by powering advanced code-generation tools such as GitHub Copilot, 
or DeepSeek-Coder~\cite{DBLP:journals/corr/abs-2107-03374,10.1109/MS.2023.3248401}. While these models significantly boost developer productivity, they also introduce serious security risks~\cite{10.1145/3510003.3510146,10431665}. 
Despite their strong empirical performance, the internal mechanisms through which LLMs recognize malicious or vulnerable code patterns remain poorly understood. This lack of interpretability raises important concerns regarding trustworthiness, robustness, and security, particularly in safety-critical software development scenarios where incorrect or manipulated predictions may introduce severe risks~\cite{10.1145/3610721,DBLP:journals/corr/abs-2307-15043}.
Such vulnerabilities stem from the models' tendency to memorize and reproduce harmful patterns present in their pre-training or fine-tuning data, especially from public 
repositories and security-related datasets~\cite{DBLP:conf/uss/CarliniTWJHLRBS21,DBLP:journals/corr/abs-2311-17035}. To mitigate this, 
current defense strategies primarily rely on prompt engineering, safety fine-tuning, 
or post-hoc filtering. However, these approaches treat the model as a black box and often fail in adaptive or 
adversarial
prompts~\cite{DBLP:conf/emnlp/PerezHSCRAGMI22,DBLP:conf/emnlp/WallaceFKGS19}. 

\vspace{.2cm}
\noindent
\textbf{Motivation.} 
Recent studies have shown that safety mechanisms and refusal behaviors are not uniformly distributed across the model but are encoded in specific internal representations~\cite{palma-etal-2025-llamas}. 
There has been evidence showing that LLMs can memorize insecure coding patterns, generate vulnerable code, and remain susceptible to adversarial manipulation~\cite{DBLP:conf/icsm/BosnakMLK25,SPINA2026112729}. 
Despite this insight, there is limited understanding of which layers are primarily responsible for detecting and blocking malicious code. 
Thus, studying which neurons and 
representations contribute to malicious code detection is crucial for improving 
transparency and enabling fine-grained control over model behavior.

\vspace{.1cm}
\noindent
\textbf{Objectives.} Our work aims to bridge such a gap by systematically probing the internal neurons of 
code 
LLMs to identify those that most effectively detect 
malicious code. Using benign and malicious PyPI packages~\cite{guo2023empirical}, we attribute malware-detection behavior to individual feed-forward neurons and verify the attribution through causal interventions. Having pinpointed where and how LLMs recognize malicious code is a prerequisite for both future safeguards and halting the malware generation of such requests.

\vspace{.1cm}
\noindent
\textbf{Methods and Findings.} Experiments on 
\Ll, \Qw, and \Ms reveal that guardrail activation is highly layer-dependent, \ie 
middle layers for \Ll and \Ms, late layers for \Qw. By localizing guardrail mechanisms, our work provides 
practical benefits: \emph{(1)} enabling lightweight, early-exit safety checks that significantly reduce inference cost while maintaining high detection accuracy; \emph{(2)} supporting targeted model editing and layer-specific unlearning to strengthen defenses without degrading general code-generation capability; and \emph{(3)} 
laying the foundation for self-monitoring defenses that let a model recognize and reject malicious generation at its source.

\vspace{.1cm}
\noindent
\textbf{Contributions.} Through this preliminary study as a NIER paper, we advocate for fundamentally new research directions in probing and understanding LLMs for software engineering tasks. 
Our work represents an initial step toward systematically investigating the 
behaviors 
of LLMs in the detection of malicious code, 
making the following contributions.


\begin{itemize}
    \item A practical approach to understand 
    LLMs' 
    behaviors in recognizing malicious code.
    \item An empirical study with three LLMs on real-world datasets to validate the performance.
    \item A replication package including code and data 
    to foster open science~\cite{Artifacts}.
\end{itemize}

\noindent
\textbf{Structure.} 
Section~\ref{sec:proposal} describes our methodology, covering dataset construction, neuron attribution methods, and intervention conditions. Section~\ref{sec:settings} details the experimental setup. Section~\ref{sec:rs} reports and discusses the results addressing our research questions. Finally, Section~\ref{sec:con} concludes the paper and outlines directions for future work.


\section{Methodology} 
\label{sec:proposal}



\begin{figure}
    \centering
    \includegraphics[width=0.98\linewidth]{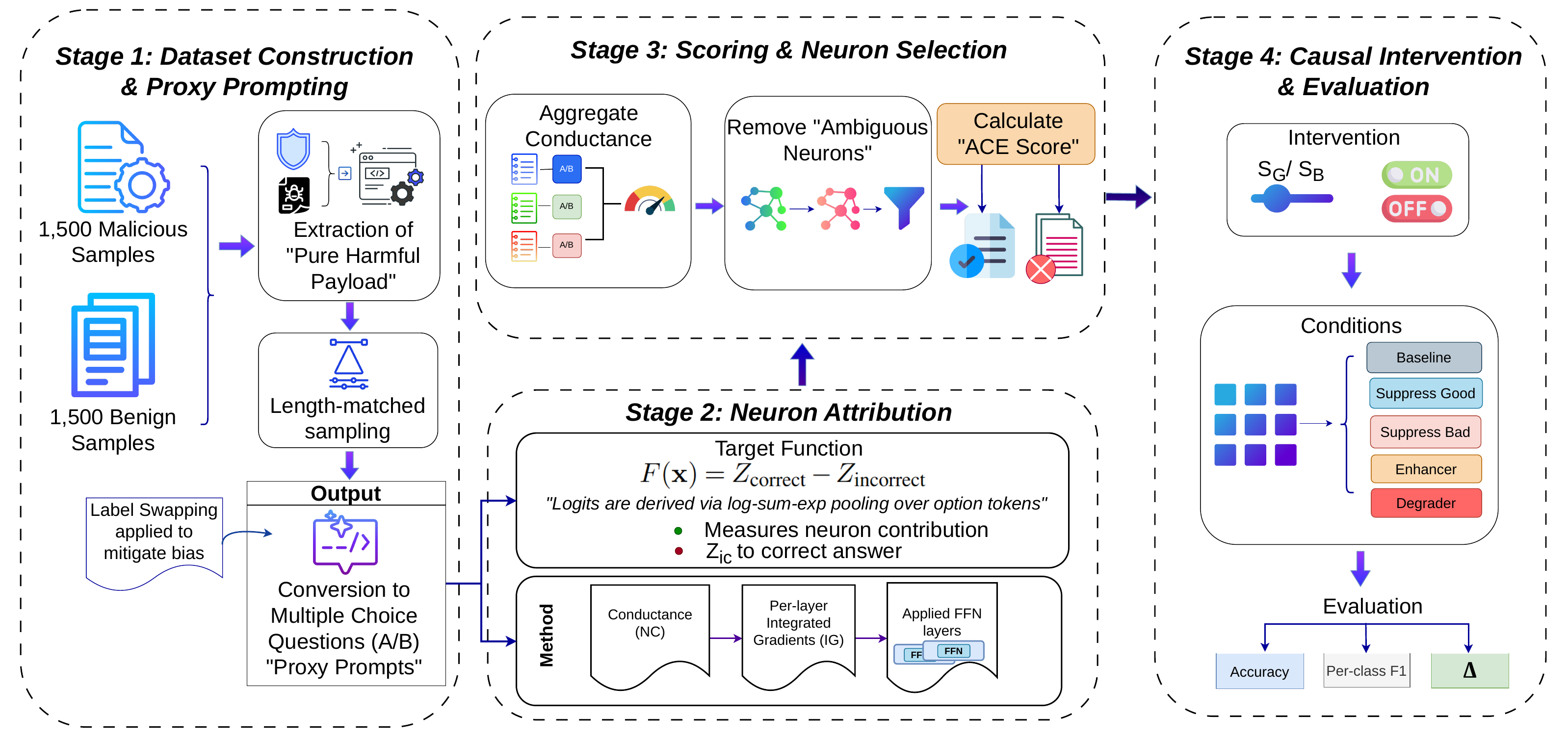}
    \caption{Framework overview: Neuron-level detection of malicious code patterns.}
    \label{fig:Pipeline}
\end{figure}


We employed various mechanistic interpretability methods~\cite{dhamdhere2018important,li2026identifyinggoodbadneurons} to pinpoint the neurons that support and hinder the model's ability to detect malicious code. 
Both malware and benign samples were collected to perform contrastive analysis of the models' internal representations when detecting malware versus benign code. 
Shown in Fig.~\ref{fig:Pipeline}, our pipeline consists of four main stages, i.e., \emph{(1)} Dataset Construction \& Proxy Prompting; \emph{(2)} Neuron Attribution; \emph{(3)} Scoring \& Neuron Selection; and \emph{(4)} Causal Intervention \& Evaluation, 
explained as follows.

\subsection{Dataset Construction and Proxy Prompting}

Malicious PyPI packages leverage Python automation mechanisms at two key stages: install-time and import-time~\cite{guo2023empirical}. 
We developed a heuristic to collect the pure harmful payload, avoiding unwanted benign implementation in those malware packages. In particular, instead of throwing all package files into the malware samples, which may lead to false positives due to the cloning implementation, we scan for the setup file (\texttt{setup.py}), which is run during installation. For runtime attacks, we look for the initial file (often named \texttt{\_\_init\_\_.py}) that is suspiciously long, which may indicate it contains the malicious payload. For benign samples, to create a suitable contrastive set with the malware, we develop a heuristic to prioritize collecting package setup code files with a similar structure and length. 

From the collected dataset (cf. Table~\ref{tab:dataset}), we ask LLMs if a code snippet is malware or benign in the form of multiple choice question A/B. To avoid the tendency of LLMs to pick the first answer or default to specific tokens~\cite{pezeshkpour-hruschka-2024-large,zheng2024large}, we designed the proxy question to swap the label of A/B and change the prompt for the proxy questions.

\begin{table}[ht]
  \centering
  \caption{Dataset distribution by line-count bucket.}
  \label{tab:dataset}
  \begin{tabular}{lrrr}
    \toprule
    \textbf{Bucket (lines)} & \textbf{Malware} & \textbf{Benign} & \textbf{Total} \\
    \midrule
    $\leq 20$       & 98 & 98 & 196 \\
    21 -- 50        & 1,107 & 1,107 & 2,214 \\
    51 -- 100       & 110 & 110 & 220 \\
    101 -- 200      & 139 & 139 & 278 \\
    $> 200$         & 46 & 46 & 92 \\
    \midrule
    \textbf{Total}  & \textbf{1,500} & \textbf{1,500} & \textbf{3,000} \\
    \bottomrule
  \end{tabular}
\end{table}

\subsection{Neuron Attribution} 

Integrated gradients (IG)
~\cite{10.5555/3305890.3306024} is an axiomatic attribution method that assigns contributions to inputs by accumulating the gradients along an interpolation path from the baseline input $x'$ to the actual input $x$ as expressed in the following formula.  
Let $F: \mathbb{R}^n \to [0, 1]$ be the function representing a deep neural network, $\mathbf{x} \in \mathbb{R}^n$ be the input, $\mathbf{x'} \in \mathbb{R}^n$ be the baseline input. The integrated gradients calculate the attribution of each input entry by integrating the gradients along the interpolation path from $x'$ to $x$.
$$\text{IntegratedGrads}_i(x) \coloneq (x_i - x'_i) \times \int_{\alpha=0}^{1} \frac{\partial F(x' + \alpha \times (x-x'))}{\partial x_i} d\alpha$$
The target function F is defined as: $F(x) = Z_{correct} - Z_{incorrect}$, in which $Z_{correct}$ is the raw logit at the output layer corresponding to the correct answer's token prior to the softmax activation; and $Z_{incorrect}$ is the raw logit value corresponding to the incorrect answer. 
By applying the target function $F(x)$ as given, we can equally reward a neuron for amplifying the correct answer, i.e., increasing $Z_{correct}$ and penaltying the wrong answer $Z_{incorrect}$.

Two methods \cite{dhamdhere2018important,li2026identifyinggoodbadneurons} are applied to identify neurons responsible for malware detection. 

\smallskip
\noindent
$\triangleright$ \textbf{Per-layer IG on FFN neurons (IG).} 
Our second attribution method builds on the framework proposed by Li et al.~\cite{li2026identifyinggoodbadneurons}. Particularly, for $w^l_i$ is the i-th neuron in a l-th layer in an FFN, 
$\hat{w}^l_i$ is the neuron activation value. The integrated gradient for neuron i in layer l is: 
$$\text{IG}(\hat{w}^l_i) = \frac{\hat{w}^l_i}{m} \sum^m_{k=1} \frac{\partial F(\frac{k}{m}\cdot\hat{w}^l_i)}{\partial \hat{w}^l_i}$$

\smallskip
\noindent
$\triangleright$ \textbf{Neuron Conductance (NC).}
The Neuron Conductance method
~\cite{dhamdhere2018important} extends Integrated Gradients to attribute a model's output to individual hidden neurons by measuring how much signal that neuron conducts from the input to the output:
$$\text{Cond}_i^y(x) = (x_i - x'_i) \cdot \int_{\alpha=0}^1 \frac{\partial F(x'+\alpha(x-x'))}{\partial y} \cdot \frac{\partial y}{\partial x_i} d\alpha$$

\subsection{Scoring and Neuron Selection} 
From the given methods and framework, we adopted them to our malware detection task. 
We calculate the Example Score by summing the conductance on all 3 questions: 
$$\text{ES}_{e_j} = \sum_{t=1}^{3} \text{Cond}(w, p_t^{(j)})$$ in which, the $p_t^{(j)}$ is the t-th prompt for the j-th sample. From the Example Scores values, we picked 2 sets of neurons for each sample: \emph{(i)} $G_j$: z neurons with the highest ES; \emph{(ii)} 
$B_j$: z neurons with the lowest ES. %
After that, we filter out the ambiguous neurons, i.e., those that are considered a Facilitator in one sample but an Inhibitor in another. Formally, a neuron $(l, i)$, the i-th neuron in the l-th layer, is considered as ambiguous if 
$$(l,i) \in \bigcup_j G_j \text{ and } (l, i) \in \bigcup_j B_j$$
Finally, we calculate the aggregated conductance score by aggregating per-example example scores, with the ambiguous neurons masked out.
$$\text{ACE}(w) = \frac{1}{N} \sum_{j=1}^N 1[w \in G_j \cup B_j] \cdot 1[w \notin Ambiguous] \cdot ES_{e_j}(w)$$
The final set of neurons is selected: \emph{(i)} $G_T$: top K neurons with the highest ACE score; and \emph{(ii)} $B_T$: top K neurons with the lowest ACE score, %
K was set to 100. 

\subsection{Causal Intervention}
After selecting the good and bad neurons using both methods, we intervened the neurons' values to test whether they were functional and had effects on the models' performance. We evaluated five intervention conditions on each model. Writing $s_G$ and $s_B$ for the multiplicative scales applied to good and bad neurons respectively, there are the following configurations: 
\textbf{Baseline}~($s_G{=}s_B{=}1$, no modification), \textbf{Suppress Good}~($s_G{=}0$, $s_B{=}1$), \textbf{Suppress Bad}~($s_G{=}1$, $s_B{=}0$), \textbf{Enhancer}~($s_G{=}2$, $s_B{=}0$), and \textbf{Degrader}~($s_G{=}0$, $s_B{=}2$)~\cite{li2026identifyinggoodbadneurons}.

\section{Experimental Setup}
\label{sec:settings}


\subsection{Evaluation Metrics}

We evaluate model performance on a held-out balanced test set under five intervention
conditions. Let $\mathit{TP}$, $\mathit{FP}$, $\mathit{TN}$, and $\mathit{FN}$ denote
true positives, false positives, true negatives, and false negatives, respectively,
where the positive class is \textit{malware}, then the metrics are defined as follows.

\smallskip
\noindent
$\triangleright$ \textbf{Per-class Precision, Recall, and F1-score.} For each class $c \in \{\text{benign},\, \text{malware}\}$:
\[
  \text{Precision}_c = \frac{TP_c}{TP_c + FP_c}
  \qquad
  \text{Recall}_c    = \frac{TP_c}{TP_c + FN_c} 
  \qquad
  F1_c = \frac{2 \cdot \text{Precision}_c \cdot \text{Recall}_c}
              {\text{Precision}_c + \text{Recall}_c} 
\]

\smallskip
\noindent
$\triangleright$ \textbf{Overall Accuracy.} Accuracy and its 95\% Wilson 
confidence interval~\cite{newcombe1998two,sokolova2009systematic,wilson1927probable} are: 
\[
  \text{Accuracy} = \frac{TP + TN}{TP + TN + FP + FN}
  \qquad
  \text{CI}_{95} = \frac{\hat{p} + \dfrac{z^2}{2n}
                         \pm z\sqrt{\dfrac{\hat{p}(1-\hat{p})}{n}
                         + \dfrac{z^2}{4n^2}}}
                        {1 + \dfrac{z^2}{n}}
\]
where $\hat{p}$ is the observed accuracy, $n$ is the test-set size, and $z = 1.96$.

\smallskip
\noindent
$\triangleright$ \textbf{McNemar's Test.} To assess whether two intervention conditions produce significantly different error patterns, we apply McNemar's test with Edwards' continuity correction \cite{Edwards_1948,McNemar_1947} (paired, $\alpha = 0.05$).
Let $b$ and $c$ denote the number of samples where condition~1 is correct and
condition~2 is wrong, and vice versa. The test statistic is:
\[
  \chi^2 = \frac{(|b - c| - 1)^2}{b + c}
  \label{eq:mcnemar}
\]
which is asymptotically distributed as $\chi^2$ with one degree of freedom under
the null hypothesis that both conditions have equal error rates.

\smallskip
\noindent
$\triangleright$ \textbf{Effectiveness Gap.} To summarize the overall utility of the identified neuron sets in a single scalar, we define the effectiveness gap as:
  $\Delta = \text{Accuracy}_{\text{Enhancer}}
         - \text{Accuracy}_{\text{Degrader}}$.
A large positive $\Delta$ indicates that the identified neurons are both sufficient to boost correct predictions (Enhancer) and necessary to maintain them (Degrader), providing strong causal evidence for their role in malware detection.

\subsection{Model Settings}

\smallskip
\noindent
$\triangleright$ 
\textbf{Model Configuration and Infrastructure.}
All models are loaded in full \texttt{bfloat16} precision \emph{without} quantization--a hard requirement, as quantization introduces numerical noise that corrupts the gradient signals needed for both IG and Conductance attribution. Models are deployed with automatic device mapping across available GPUs. \Ll consists of 8B parameters with 32 layers, while \Ms and \Qw both have 7B parameters, with 32 and 28 layers, respectively.

\smallskip
\noindent
$\triangleright$ 
\textbf{Attribution Hyperparameters.}
For IG, the $m = 16$ integration steps are approximated via a right Riemann sum 
(excluding $\alpha = 0$); since IG patches one layer at a time, the total forward-pass cost per sample is $m \times L$ (e.g., $16 \times 32 = 512$ for 32-layer models). Conductance 
interpolates input embeddings with the same step budget but runs all layers freely, requiring only $m$ forward-backward passes total - substantially cheaper than per-layer IG. The attribution sample count $|\mathcal{D}_\text{attrib}|$ is swept over $\{10, 20, 50, 100, 200\}$ to assess sensitivity; $20$ is the default used in all primary results.

\smallskip
\noindent
$\triangleright$ 
\textbf{Prompting and Truncation.}
Source code samples are truncated to a maximum of 2,000 
characters; samples exceeding this limit receive a 
\texttt{\textbackslash n...[truncated]} suffix. The truncated 
code is embedded in a structured binary-choice prompt 
(Section~\ref{sec:proposal}). Model predictions are derived 
from logit argmax over the space-prefixed token sets 
$\{\texttt{" A"},\,\texttt{" B"}\}$, aggregated via 
log-sum-exp pooling to handle multiple tokenization variants of each option letter. This 
strategy is applied identically at both attribution and evaluation time, ensuring that the target function $F$ optimized during attribution matches the criterion used to measure accuracy.

\smallskip
\noindent
$\triangleright$ 
\textbf{Data Split.}
The length-matched dataset 
was partitioned
by both class label and line-count bucket, yielding a held out test set of 2,890 samples (1,445 benign, 1,445 malware) that is 
used exclusively for intervention evaluation and is separated from the 110 samples used for neuron attribution and ACE scoring.

\section{Results and Discussion}
\label{sec:rs}

\subsection{\rqone}

Causal interventions on the identified neurons produce measurable shifts in accuracy and per-class F1, 
with effect sizes differing substantially by model and method.


\begin{table}[h!]
\centering
\footnotesize
\caption{Intervention accuracy and Effectiveness Gap $\Delta$. 95\% Wilson
CIs are within $\pm 0.019$ of every point estimate (tightest, $\pm 0.009$,
at the near-ceiling rows). ``n.s.''~=~not
significant at $\alpha=0.05$; bold $p$-values are significant.}
\label{tab:intervention_summary}
\small
\footnotesize
\tiny
\begin{tabular}{llccccccc}
\toprule
\textbf{Model} & \textbf{Method} & \textbf{Base} & \textbf{Sup-G} & \textbf{Sup-B} & \textbf{Enh} & \textbf{Deg} & $\boldsymbol{\Delta}$ & \textbf{McNemar} \\
\midrule
\Ll   & IG & 0.939 & 0.513 & 0.939 & 0.941 & 0.500 & 0.441 & 0.386 n.s. \\
\Ll   & NC & 0.939 & 0.901 & 0.932 & 0.941 & 0.500 & 0.441 & 0.359 n.s. \\
\Ms & IG & 0.665 & 0.625 & 0.920 & 0.941 & 0.616 & 0.325 & \textbf{<0.001} \\
\Ms & NC & 0.665 & 0.652 & 0.696 & 0.800 & 0.645 & 0.154 & \textbf{<0.001} \\
\Qw    & IG & 0.939 & 0.939 & 0.940 & 0.941 & 0.832 & 0.109 & 0.181 n.s. \\
\Qw    & NC & 0.939 & 0.941 & 0.939 & 0.939 & 0.932 & 0.007 & 0.617 n.s. \\
\bottomrule
\end{tabular}
\end{table}

\smallskip
\noindent
$\triangleright$ \textbf{Intervention effects on accuracy.} 
Effect sizes varied across models, i.e., the high-baseline models \Ll and \Qw (93.9\%) show little response to corrective interventions (Suppress Bad, Enhancer), whereas \Ms improves from 66.5\% to 94.1\% under the IG Enhancer. IG Suppress Bad alone raises it to 92.0\%, suggesting that these inhibitory neurons drive \Ms's low baseline. NC shows the same direction but weaker, with its Enhancer reaching only 80\%.

Under destructive interventions (Suppress Good, Degrader), \Ll and \Qw diverge: the former 
collapses to 50.0\% under the Degrader (both methods) and 51.3\% under IG Suppress Good, proving its identified neurons are causally necessary. The latter 
resists with only the IG Degrader has a mild effect (83.2\%). \Ms degrades slightly, bounded by its already-low baseline.

IG and NC agreed on direction but not on magnitude: identical on \Ll's Degrader ($\Delta=0.441$ for both) but divergent sharply on Suppress Good (IG: 51.3\% and NC: 90.1\%). The effectiveness gap $\Delta_\text{IG}/\Delta_\text{NC}$ is roughly 2 times on \Ms (0.325 vs 0.154) and over 10 times on \Qw (0.109 vs 0.007). IG produces a gap of equal size or larger in every row of the table.


\begin{figure}[h!]
    \centering
    \includegraphics[width=0.95\linewidth]{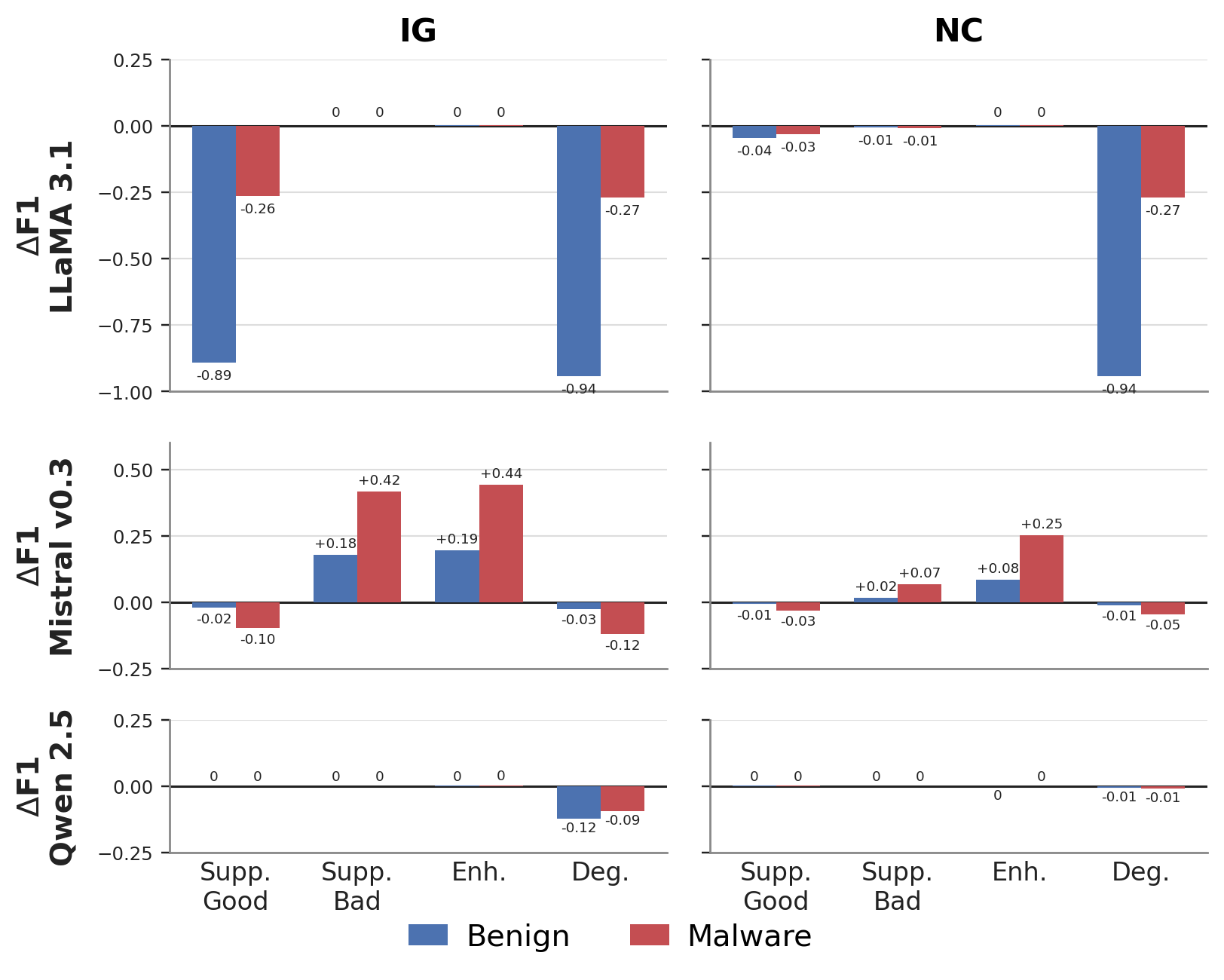}
    \caption{Per-class $\Delta$F1 relative to baseline for each 
    combination. Blue bars = benign class; red bars = malware class. Positive values = improvement; negative values = degradation.}
    \label{fig:f1_delta}
\end{figure}
\smallskip
\noindent
$\triangleright$ \textbf{Intervention effects on Per-class F1.} As shown in Fig.~\ref{fig:f1_delta}, 
the models behave differently to the interventions. Those with strong initial baselines, such as \Ll and \Qw, respond minimally to Suppress Bad and Enhancer, both yield $\Delta\text{F1} \approx0$. \Ms, whose baseline was asymmetric (benign 0.99, malware recall 0.33), responds more clearly: IG Suppress Bad and the Enhancer raise malware recall to 0.84–0.89 while holding benign recall at 0.99. 

The destructive interventions, such as Suppress Good and Degrader, separate \Ll and \Qw, both of which were low responders to the corrective interventions. \Ll collapses under both IG and NC Degrader ($\Delta\text{F1}_\text{benign} =-0.94$, $\Delta\text{F1}_\text{malware}=-0.27$) and IG Suppress Good ($\Delta\text{F1}_\text{benign} =-0.89$, $\Delta\text{F1}_\text{malware}=-0.26$), whereas \Qw stays stable with only slight degradation under IG Degrader. \Ms also degrades slightly with its $\Delta\text{F1}_\text{benign} =-0.03$, $\Delta\text{F1}_\text{malware}=-0.12$ under the IG Degrader.
NC showed less visible effects than IG overall: dramatic only on \Ll's Degrader, and weaker across other models and intervention types.

\vspace{.2cm}
\begin{tcolorbox}[boxrule=0.86pt,left=0.3em, right=0.3em,top=0.1em, bottom=0.05em]
	\small{\textbf{Answer to RQ$_1$.} Both IG and NC identify neurons that causally encode malware-detection behavior, though effect sizes are strongly model-dependent. \Ms suppresses inhibiting neurons alone increases the accuracy. 
        For \Ll and \Qw, 
        the Enhancer yields negligible gains, yet the Degrader collapses \Ll to 50.0\%. Meanwhile, \Qw is relatively robust and stable to these intervention effects. 
        Overall, IG achieves larger or equal effectiveness gaps compared to NC.}
\end{tcolorbox}

\subsection{\rqtwo}

We investigate where in the network these neurons reside and whether attribution consistently points to the same regions across different data subsets.


\begin{figure} [h!]
    \centering
    \includegraphics[width=0.95\linewidth]{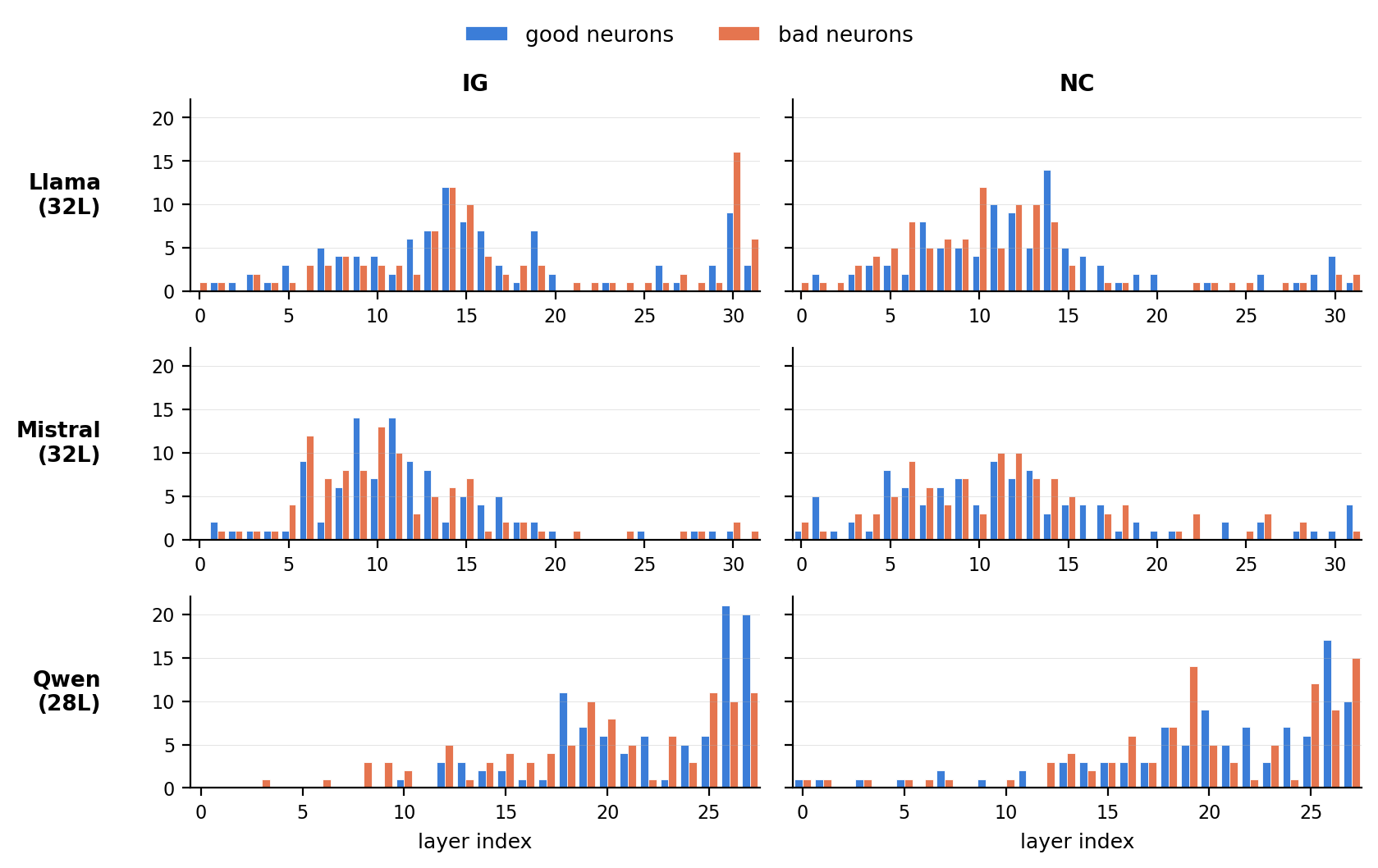}
    \caption{Per-layer distribution of top-100 good (blue) and bad (orange) neurons. 
    }
    \label{fig:layer_dist}
\end{figure}

\smallskip
\noindent
$\triangleright$ \textbf{Distribution.} The results are shown in Fig.~\ref{fig:layer_dist}.
\Ll (32 layers): Both IG and NC place the primary concentration in the middle layers. Under IG, good neurons peak in Layer 13, and bad neurons peak in Layers 14-15, with an additional spike in Layer 30. NC shows a similar mid-layer peak at Layer 14 with good and bad neurons co-located throughout Layers 6-18, plus a modest late spike at Layer 30. 

\Ms (32 layers): The concentration is the tightest among all three models. Both IG and Conductance place good and bad neurons almost exclusively in layers 5-14, with IG peaking at layer 10 and Conductance still within the same band. After layer 15, both methods select almost no neurons, and the IG and Conductance distributions are most similar across all three models.

\Qw (28 layers): The dominant signal is in the late layers. Using IG, the distribution shows a secondary cluster around layer 12, followed by a sharp spike at layers 25-26, peaking at layer 26, the highest single-layer count across all models. Under NC, the signal is more spread across layers 10-27, with the same peak at layer 26 but less abrupt than IG. The first ten layers carry almost no signal under either method.

Each model exhibits a clear preferred layer region: middle layers for \Ll and \Ms, late layers for \Qw. \emph{A consistent pattern across all three models is that good and bad neurons co-localize in the same layer bands rather than being separated into different regions of the network.}

\smallskip
\noindent
$\triangleright$ \textbf{Stability.} We test whether the neuron selection is consistent by repeating attribution five times on different random subsets of the data (K = 100 neurons per run, seeds 0-4). 
We compute two complementary similarities over the 10 pairs of runs: Jaccard similarity of the selected neuron sets, capturing whether the same individual neurons are picked; and cosine similarity of the per-layer neuron-count histograms. 

\begin{figure}[h!]
    \centering
    \includegraphics[width=0.95\linewidth]{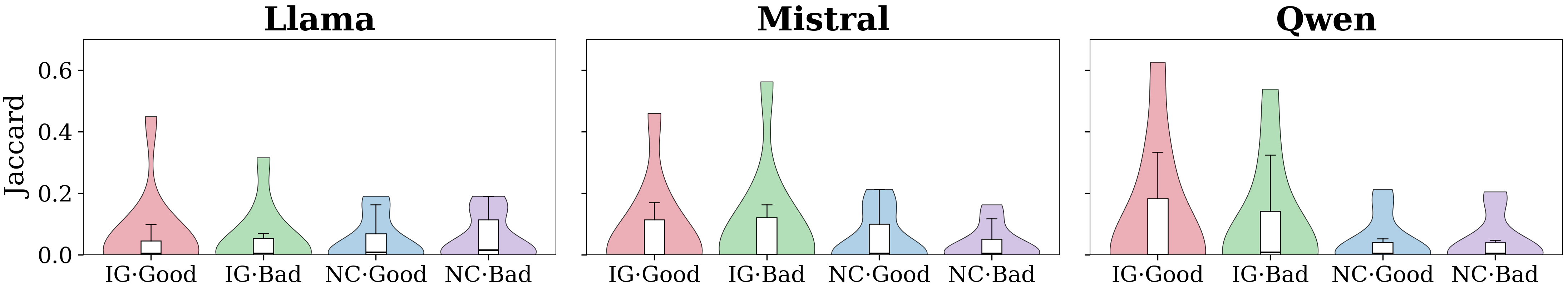}
    \caption{Pairwise Jaccard similarity of the top-100 neuron sets across 5 stability runs.} 
    \label{fig:stability_jaccard}
\end{figure}
As shown in Fig.~\ref{fig:stability_jaccard}, Jaccard values are low across all models and methods (all medians below 0.20), showing that the specific neurons identified do not reliably repeat across attribution subsets. \Ll IG is the least consistent (medians $\approx$0.03-0.04 for both classes, upper tails to $\approx$0.45), with NC slightly more stable ($\approx$0.10). \Ms IG has higher medians ($\approx$0.10 good, $\approx$0.12 bad) and a wide upper tail for bad neurons ($\approx$0.57). \Qw IG-Good has the highest median of all conditions ($\approx$0.17, tail to 0.62), reflecting the sharp single-layer concentration in Fig.~\ref{fig:layer_dist}, whereas \Qw NC collapses to near-zero ($\approx$0.03).

\begin{figure}[h!]
    \centering
    \includegraphics[width=0.95\linewidth]{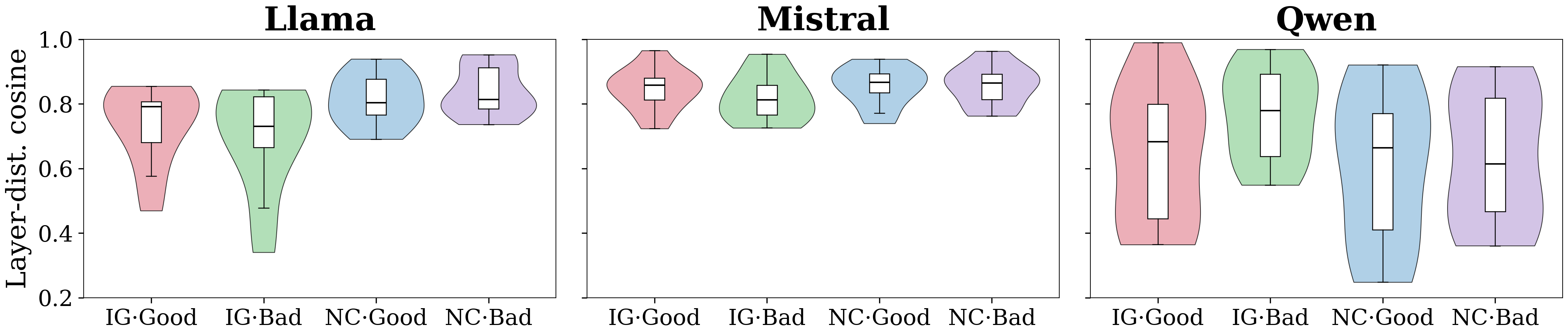}
    \caption{Pairwise cosine similarity of the per-layer neuron-count histograms across the same 5 runs. 
    Medians span between 0.6 and 0.9, an order of magnitude above the corresponding Jaccards.}
    \label{fig:stability_cosine}
\end{figure}

As shown in Fig.~\ref{fig:stability_cosine},  
despite low Jaccard values, the layer-level shape of the selection is substantially more stable: cosine values across all conditions range from 0.61 to 0.86, an order of magnitude above the corresponding Jaccard. 
\Ms is the most stable at the layer level. All four conditions (IG/NC $\times$ Good/Bad) have medians of 0.82-0.86 with narrow IQRs ($\approx$0.78-0.90), so the early-to-middle band (layers 5-14) is consistently identified regardless of which neurons within it are selected. \Ll is intermediate, with NC more stable than IG: NC-Good ($0.813 \pm 0.076$) and NC-Bad ($0.840 \pm 0.073$) are tighter than IG-Good ($0.732 \pm 0.120$) and IG-Bad ($0.694 \pm 0.159$). IG-Bad shows the widest spread (outliers to $\approx$0.45) because the late layer-30 spike is not always selected. 
\Qw is the least stable, having the widest IQRs of all three models (especially NC-Good, $0.606 \pm 0.231$, run pairs ranging cosine $\approx$0.4-0.8): IG concentrates sharply at layers 25-26 while NC spreads across the final third. 


\vspace{.2cm}
\begin{tcolorbox}[boxrule=0.86pt,left=0.3em, right=0.3em,top=0.1em, bottom=0.05em]
	\small{\textbf{Answer to RQ$_2$.} Each model concentrates its malware-detection signal in a unique layer band: middle layers for \Ll and \Ms, late layers for \Qw. The selected neurons are not stable across runs. However, stability resides at the layer level. \Ll and \Ms, which exhibit this layer stability behavior, also respond most clearly to causal intervention effects in RQ1, whereas \Qw, whose layer distribution is the most dispersed, responds the least, a co-occurrence suggesting that stable layer-level localization may be a prerequisite for effective causal interventions.}
\end{tcolorbox}


\subsection{Discussion}

\noindent
$\triangleright$ 
\textbf{Implications.} 
By probing the neurons associated with security-relevant knowledge, we can gain deeper insights into how LLMs internally encode malicious programming concepts and insecure coding patterns. Such understanding enables the identification of potentially harmful memorized behaviors and paves the way toward more reliable defense mechanisms, including neuron-level editing, selective machine unlearning, and security-aware alignment for code-focused LLMs. Moreover, this line of research opens opportunities for fine-grained security interventions at the neuron and layer levels, such as targeted suppression of harmful behaviors, activation steering, and layer-specific hardening strategies. For instance, understanding which layers are more strongly associated with malicious code recognition may help explain why certain adversarial attacks succeed, e.g., by bypassing safety-relevant intermediate representations, and may inspire the integration of dedicated safety adapters. 

\vspace{.1cm}
\noindent
$\triangleright$ 
\textbf{Limitations.} 
Due to limited computational 
resources, we were able to conduct experiments with relatively small LLMs, \ie less than 8B parameters. Being a NIER 
contribution, \emph{the primary objective of our work is to introduce and demonstrate the feasibility of a novel probing-based approach for understanding malicious code detection in LLMs, rather than to provide a large-scale or exhaustive evaluation.} %
Essentially, since the proposed probing  methodology does not require expensive retraining or architecture modifications, it can be scaled to larger and more capable code-generation 
models with a lot 
more layers in a relatively lightweight and model-agnostic manner. 
We believe that the encouraging preliminary results motivate further investigation and broader 
empirical validation in future work.


\subsection{Threats to Validity}

\noindent
$\triangleright$ \textbf{Internal Validity.} Attribution identifies a layer-level signal that is stable across seeds, but the specific neurons selected are not: pairwise Jaccard similarity remains below 0.20 (Fig.~\ref{fig:stability_jaccard}). Our conclusions, therefore, hold at the granularity of layer bands rather than at the level of individual neurons. Attribution hyperparameters ($m{=}16$, $K{=}100$,
$|\mathcal{D}_\text{attrib}|{=}20$) are also fixed. Hence, a sensitivity analysis is left to future work.

\noindent
$\triangleright$ \textbf{External Validity.} Our findings are derived exclusively from three 
7-8B LLMs 
and Python/PyPI packages, focusing on install-time and import-time payloads. They may not transfer to larger or differently aligned models, to other ecosystems (e.g., npm, Maven) or attack vectors (e.g., typosquatting), or to malware more recent than the PyPI Malregistry~\cite{guo2023empirical}.



\section{Conclusion and Future Work}
\label{sec:con}


This paper proposes a practical approach to locate the FFN neurons responsible for distinguishing malicious from benign code inside three instruction-tuned LLMs, i.e., \Ll, \Ms, \Qw. Using two attribution methods, we attributed the detection behavior to a specific set of neurons and verified the attribution using causal interventions.
Furthermore, we found that each model concentrates its signal in a distinct layer band, and that neuron localization is stable at the layer level rather than at the individual neuron level.
This work provides a foundation for LLM safety behavior in recognizing malicious code, offers open avenues for early-exit safety checks, and enables targeted model editing. 
For future work, we plan to bridge code recognition and generative refusal, disentangle surface-level cues from semantic intent, and extend the analysis to larger models and other software ecosystems.


\section{Data Availability}
The replication package of our work is available in an anonymized public repository~\cite{Artifacts}. It includes the source code, data, and instructions in the form of a detailed README.MD file for reproducing the results obtained 
in this paper.

\section{Acknowledgments}
This paper has been partially supported by the MOSAICO project (Management, Orchestration and Supervision of AI-agent COmmunities for reliable AI in software engineering) that has received funding from the European Union under the Horizon Research and Innovation Action (Grant Agreement No. 101189664). 




\bibliography{main}

@ARTICLE{10431665,
  author={Yang, Zhou and Xu, Bowen and Zhang, Jie M. and Kang, Hong Jin and Shi, Jieke and He, Junda and Lo, David},
  journal={IEEE Transactions on Software Engineering}, 
  title={Stealthy Backdoor Attack for Code Models}, 
  year={2024},
  volume={50},
  number={4},
  pages={721-741},
  doi={10.1109/TSE.2024.3361661}}

@inproceedings{10.1145/3510003.3510146,
author = {Yang, Zhou and Shi, Jieke and He, Junda and Lo, David},
title = {Natural attack for pre-trained models of code},
year = {2022},
isbn = {9781450392211},
publisher = {Association for Computing Machinery},
address = {New York, NY, USA},
url = {https://doi.org/10.1145/3510003.3510146},
doi = {10.1145/3510003.3510146},
booktitle = {Proceedings of the 44th International Conference on Software Engineering},
pages = {1482–1493},
numpages = {12},
location = {Pittsburgh, Pennsylvania},
series = {ICSE '22}
}

@article{10.1109/MS.2023.3248401,
author = {Ozkaya, Ipek},
editor = {Ipek Ozkaya},
title = {Application of Large Language Models to Software Engineering Tasks: Opportunities, Risks, and Implications},
year = {2023},
issue_date = {May-June 2023},
publisher = {IEEE Computer Society Press},
address = {Washington, DC, USA},
volume = {40},
number = {3},
issn = {0740-7459},
url = {https://doi.org/10.1109/MS.2023.3248401},
doi = {10.1109/MS.2023.3248401},
journal = {IEEE Softw.},
month = may,
pages = {4–8},
numpages = {5}
}

@inproceedings{DBLP:conf/icsm/BosnakMLK25,
  author       = {Ahmet Emir Bosnak and
                  Sahand Moslemi and
                  Mayasah Lami and
                  Anil Koyuncu},
  title        = {Explicit Vulnerability Generation with LLMs: An Investigation Beyond
                  Adversarial Attacks},
  booktitle    = {{IEEE} International Conference on Software Maintenance and Evolution,
                  {ICSME} 2025, Auckland, New Zealand, September 7-12, 2025},
  pages        = {821--826},
  publisher    = {{IEEE}},
  year         = {2025},
  url          = {https://doi.org/10.1109/ICSME64153.2025.00086},
  doi          = {10.1109/ICSME64153.2025.00086},
  bibsource    = {dblp computer science bibliography, https://dblp.org}
}

@article{SPINA2026112729,
title = {Peeking inside the black box: Training data exposure in code language models},
journal = {Journal of Systems and Software},
volume = {234},
pages = {112729},
year = {2026},
issn = {0164-1212},
doi = {https://doi.org/10.1016/j.jss.2025.112729},
url = {https://www.sciencedirect.com/science/article/pii/S016412122500398X},
author = {Angelica Spina and Marco Russodivito and Simone Scalabrino and Rocco Oliveto}
}

@misc{Artifacts,
	author       = {{Anonymous}},
	title        = {{Replication Package for ``Which Neurons Detect Malicious Code? A Probing Study of LLM Security Knowledge''}},
	year         = {2026},
	howpublished = {\url{https://anonymous.4open.science/r/esem-neuron-package/}},
	note         = {Accessed: 2026-05-27}
}

@article{DBLP:journals/corr/abs-2311-17035,
  author       = {Milad Nasr and
                  Nicholas Carlini and
                  others},
  title        = {Scalable Extraction of Training Data from (Production) Language Models},
  journal      = {CoRR},
  volume       = {abs/2311.17035},
  year         = {2023},
  url          = {https://doi.org/10.48550/arXiv.2311.17035},
  doi          = {10.48550/ARXIV.2311.17035},
  eprinttype   = {arXiv},
  eprint       = {2311.17035},
  bibsource    = {dblp computer science bibliography, https://dblp.org}
}

@inproceedings{DBLP:conf/emnlp/WallaceFKGS19,
  author       = {Eric Wallace and
                  Shi Feng and
                  Nikhil Kandpal and
                  Matt Gardner and
                  Sameer Singh},
  editor       = {Kentaro Inui and
                  Jing Jiang and
                  Vincent Ng and
                  Xiaojun Wan},
  title        = {Universal Adversarial Triggers for Attacking and Analyzing {NLP}},
  booktitle    = {Proceedings of the 2019 Conference on Empirical Methods in Natural
                  Language Processing and the 9th International Joint Conference on
                  Natural Language Processing, {EMNLP-IJCNLP} 2019, Hong Kong, China,
                  November 3-7, 2019},
  pages        = {2153--2162},
  publisher    = {Association for Computational Linguistics},
  year         = {2019},
  url          = {https://doi.org/10.18653/v1/D19-1221},
  doi          = {10.18653/V1/D19-1221},
  bibsource    = {dblp computer science bibliography, https://dblp.org}
}

@inproceedings{DBLP:conf/emnlp/PerezHSCRAGMI22,
  author       = {Ethan Perez and
                  Saffron Huang and
                  H. Francis Song and
                  Trevor Cai and
                  Roman Ring and
                  John Aslanides and
                  Amelia Glaese and
                  Nat McAleese and
                  Geoffrey Irving},
  editor       = {Yoav Goldberg and
                  Zornitsa Kozareva and
                  Yue Zhang},
  title        = {Red Teaming Language Models with Language Models},
  booktitle    = {Proceedings of the 2022 Conference on Empirical Methods in Natural
                  Language Processing, {EMNLP} 2022, Abu Dhabi, United Arab Emirates,
                  December 7-11, 2022},
  pages        = {3419--3448},
  publisher    = {Association for Computational Linguistics},
  year         = {2022},
  url          = {https://doi.org/10.18653/v1/2022.emnlp-main.225},
  doi          = {10.18653/V1/2022.EMNLP-MAIN.225},
  bibsource    = {dblp computer science bibliography, https://dblp.org}
}

@inproceedings{DBLP:conf/uss/CarliniTWJHLRBS21,
  author       = {Nicholas Carlini and
                  Florian Tram{\`{e}}r and
                  others},
  title        = {Extracting Training Data from Large Language Models},
  booktitle    = {30th {USENIX} Security Symposium, {USENIX} Security 2021, August 11-13,
                  2021},
  pages        = {2633--2650},
  publisher    = {{USENIX} Association},
  year         = {2021},
  url          = {https://www.usenix.org/conference/usenixsecurity21/presentation/carlini-extracting},
  bibsource    = {dblp computer science bibliography, https://dblp.org}
}

@article{DBLP:journals/corr/abs-2307-15043,
  author       = {Andy Zou and
                  Zifan Wang and
                  J. Zico Kolter and
                  Matt Fredrikson},
  title        = {Universal and Transferable Adversarial Attacks on Aligned Language
                  Models},
  journal      = {CoRR},
  volume       = {abs/2307.15043},
  year         = {2023},
  url          = {https://doi.org/10.48550/arXiv.2307.15043},
  doi          = {10.48550/ARXIV.2307.15043},
  eprinttype   = {arXiv},
  eprint       = {2307.15043},
  bibsource    = {dblp computer science bibliography, https://dblp.org}
}

@article{10.1145/3610721,
author = {Pearce, Hammond and Ahmad, Baleegh and Tan, Benjamin and Dolan-Gavitt, Brendan and Karri, Ramesh},
title = {Asleep at the Keyboard? Assessing the Security of GitHub Copilot’s Code Contributions},
year = {2025},
issue_date = {February 2025},
publisher = {Association for Computing Machinery},
address = {New York, NY, USA},
volume = {68},
number = {2},
issn = {0001-0782},
url = {https://doi.org/10.1145/3610721},
doi = {10.1145/3610721},
journal = {Commun. ACM},
month = jan,
pages = {96–105},
numpages = {10}
}

@article{DBLP:journals/corr/abs-2107-03374,
  author       = {Mark Chen and
                  Jerry Tworek and
                  others},
  title        = {Evaluating Large Language Models Trained on Code},
  journal      = {CoRR},
  volume       = {abs/2107.03374},
  year         = {2021},
  url          = {https://arxiv.org/abs/2107.03374},
  eprinttype   = {arXiv},
  eprint       = {2107.03374},
  bibsource    = {dblp computer science bibliography, https://dblp.org}
}

@misc{li2026identifyinggoodbadneurons,
      title={Identifying Good and Bad Neurons for Task-Level Controllable LLMs}, 
      author={Wenjie Li and Guansong Pang and Hezhe Qiao and Debin Gao and David Lo},
      year={2026},
      eprint={2601.04548},
      archivePrefix={arXiv},
      primaryClass={cs.CL},
      url={https://arxiv.org/abs/2601.04548}, 
}

@inproceedings{palma-etal-2025-llamas,
    title = "{LL}a{MA}s Have Feelings Too: Unveiling Sentiment and Emotion Representations in {LL}a{MA} Models Through Probing",
    author = "Di Palma, Dario  and
      De Bellis, Alessandro  and
      Servedio, Giovanni  and
      Anelli, Vito Walter  and
      Narducci, Fedelucio  and
      Di Noia, Tommaso",
    editor = "Che, Wanxiang  and
      Nabende, Joyce  and
      Shutova, Ekaterina  and
      Pilehvar, Mohammad Taher",
    booktitle = "Proceedings of the 63rd Annual Meeting of the Association for Computational Linguistics (Volume 1: Long Papers)",
    month = jul,
    year = "2025",
    address = "Vienna, Austria",
    publisher = "Association for Computational Linguistics",
    url = "https://aclanthology.org/2025.acl-long.306/",
    doi = "10.18653/v1/2025.acl-long.306",
    pages = "6124--6142",
    ISBN = "979-8-89176-251-0"
}

@inproceedings{10.5555/3305890.3306024,
author = {Sundararajan, Mukund and Taly, Ankur and Yan, Qiqi},
title = {Axiomatic attribution for deep networks},
year = {2017},
publisher = {JMLR.org},
booktitle = {Proceedings of the 34th International Conference on Machine Learning - Volume 70},
pages = {3319–3328},
numpages = {10},
location = {Sydney, NSW, Australia},
series = {ICML'17}
}

@inproceedings{guo2023empirical,
  title={An empirical study of malicious code in pypi ecosystem},
  author={Guo, Wenbo and Xu, Zhengzi and Liu, Chengwei and Huang, Cheng and Fang, Yong and Liu, Yang},
  booktitle={2023 38th IEEE/ACM International Conference on Automated Software Engineering (ASE)},
  pages={166--177},
  year={2023},
  organization={IEEE}
}

@article{dhamdhere2018important,
  author       = {Kedar Dhamdhere and
                  Mukund Sundararajan and
                  Qiqi Yan},
  title        = {How Important Is a Neuron?},
  journal      = {CoRR},
  volume       = {abs/1805.12233},
  year         = {2018},
  url          = {http://arxiv.org/abs/1805.12233},
  eprinttype   = {arXiv},
  eprint       = {1805.12233},
  bibsource    = {dblp computer science bibliography, https://dblp.org}
}

@inproceedings{
zheng2024large,
title={Large Language Models Are Not Robust Multiple Choice Selectors},
author={Chujie Zheng and Hao Zhou and Fandong Meng and Jie Zhou and Minlie Huang},
booktitle={The Twelfth International Conference on Learning Representations},
year={2024},
url={https://openreview.net/forum?id=shr9PXz7T0}
}

@inproceedings{pezeshkpour-hruschka-2024-large,
    title = "Large Language Models Sensitivity to The Order of Options in Multiple-Choice Questions",
    author = "Pezeshkpour, Pouya  and
      Hruschka, Estevam",
    editor = "Duh, Kevin  and
      Gomez, Helena  and
      Bethard, Steven",
    booktitle = "Findings of the Association for Computational Linguistics: NAACL 2024",
    month = jun,
    year = "2024",
    address = "Mexico City, Mexico",
    publisher = "Association for Computational Linguistics",
    url = "https://aclanthology.org/2024.findings-naacl.130/",
    doi = "10.18653/v1/2024.findings-naacl.130",
    pages = "2006--2017"
}

@article{sokolova2009systematic,
  title={A systematic analysis of performance measures for classification tasks},
  author={Sokolova, Marina and Lapalme, Guy},
  journal={Information processing \& management},
  volume={45},
  number={4},
  pages={427--437},
  year={2009},
  publisher={Elsevier}
}

@article{wilson1927probable,
  title={Probable inference, the law of succession, and statistical inference},
  author={Wilson, Edwin B},
  journal={Journal of the American Statistical Association},
  volume={22},
  number={158},
  pages={209--212},
  year={1927},
  publisher={Taylor \& Francis}
}

@article{newcombe1998two,
  title={Two-sided confidence intervals for the single proportion: comparison of seven methods},
  author={Newcombe, Robert G},
  journal={Statistics in medicine},
  volume={17},
  number={8},
  pages={857--872},
  year={1998},
  publisher={Wiley Online Library}
}

@article{Edwards_1948, title={Note on the “Correction for Continuity” in Testing the Significance of the Difference between Correlated Proportions}, volume={13}, DOI={10.1007/BF02289261}, number={3}, journal={Psychometrika}, author={Edwards, Allen L.}, year={1948}, pages={185–187}}

@article{McNemar_1947, title={Note on the Sampling Error of the Difference Between Correlated Proportions or Percentages}, volume={12}, DOI={10.1007/BF02295996}, number={2}, journal={Psychometrika}, author={McNemar, Quinn}, year={1947}, pages={153–157}}

@String{Computing = "Computing" }

@String{Computer = "{IEEE} Computer" }

@String{Psychometrika = "Psychometrika" }

@ArtifactSoftware{R,
    title = {R: A Language and Environment for Statistical Computing},
    author = {{R Core Team}},
    organization = {R Foundation for Statistical Computing},
    address = {Vienna, Austria},
    year = {2019},
    url = {https://www.R-project.org/},
}

\appendix

\end{document}